\documentclass[conference]{IEEEtran}
\IEEEoverridecommandlockouts
\pdfoutput=1
\usepackage{cite}
\usepackage{amsmath,amssymb,amsfonts}
\usepackage{algorithmic}
\usepackage{graphicx}
\usepackage{textcomp}
\usepackage{booktabs}
\usepackage{multirow}
\usepackage{xcolor}
\usepackage[T1]{fontenc}
\usepackage{newtxtext,newtxmath}
\usepackage{caption}
\usepackage{subcaption}
\def\BibTeX{{\rm B\kern-.05em{\sc i\kern-.025em b}\kern-.08em
    T\kern-.1667em\lower.7ex\hbox{E}\kern-.125emX}}

\begin{document}

\title{SPA-MAE: A Physics-Guided CSI Foundation Model for Wireless Physical Layer}

\author{
\IEEEauthorblockN{
Chen Chen, Weijie Jin, Hengtao He, Xiaoheng Sun, Shi Jin
}
\IEEEauthorblockA{
School of Information Science and Engineering, Southeast University, Nanjing 210096, China\\
Email: \{chen\_chen, jinweijie, hehengtao, sunxiaoheng, jinshi\}@seu.edu.cn
}
}
\maketitle

\begin{abstract}
Deep learning (DL) has been widely used in future 6G physical layer communications, but task-specific DL models are difficult to generalize across different physical layer tasks. Recently emerging wireless foundation models demonstrate strong generalization capability. However, existing methods mainly adapt pretrained language/vision models or rely on CSI reconstruction objectives for pretraining, with limited use of channel knowledge, and thus have limited performance. To address this limitation, we propose SPA-MAE, a physics-guided wireless foundation model by exploiting the adapted MAE backbone and channel knowledge. A physical prior module is developed to provide two complementary guidance signals in the pretraining stage. Specifically, the parameter-aware guidance branch extracts features from explicit multipath parameters and encourages the encoder output to align them, while the structure-aware guidance branch encourages the encoder to capture the sparse transformed-domain CSI structure obtained after a 2D FFT. After end-to-end learning, the MAE encoder will be retained for downstream tasks. Experiments on four wireless tasks show that SPA-MAE outperforms state-of-the-art CSI foundation models with smaller number of parameters, especially under low-SNR and limited-data conditions.
\end{abstract}

\begin{IEEEkeywords}
foundation model, masked autoencoder, physics-guided pretraining, wireless physical layer
\end{IEEEkeywords}

\section{Introduction}

Future 6G wireless systems are expected to meet increasingly stringent physical-layer requirements in spectral efficiency, latency, reliability, coverage, and sensing/positioning capability under highly dynamic propagation environments \cite{b11,b12}. To meet these requirements, deep learning (DL) has been widely explored for wireless physical-layer tasks \cite{b15,b16,b20}. However, 6G involve diverse propagation conditions, heterogeneous system configurations, and multiple task objectives. Task-specific DL models usually need separate designs and training for different scenarios or functions, limiting their scalability and transferability. In natural language processing and computer vision, large-scale pretraining has shown that reusable representations can be learned from massive data and efficiently adapted to diverse downstream tasks \cite{b17,b18,b19}. This paradigm motivates wireless foundation models for general-purpose physical-layer representation learning.

Recent studies have explored wireless foundation models from two main directions. The first direction is to adapt large AI models pretrained on non-wireless data to wireless tasks. For example, LVM4CSI exploits the structural similarity between channel state information (CSI) and visual data, and applies frozen large vision models with lightweight task layers to CSI-related tasks \cite{b4}. LLM4WM adapts a pretrained language model to multiple channel-associated tasks through wireless-specific interfaces and efficient multi-task fine-tuning \cite{b10}. These studies address the limitations of task-specific DL networks and demonstrate the potential of adapting large AI models. The second direction is to build foundation models for wireless communications directly from channel data. LWM was first proposed to learn task-agnostic channel embeddings through masked channel modeling, enabling the extraction of universal and rich features in complex environments \cite{b1}. Then, CSI-MAE was proposed to further expand the range of downstream tasks, including sensing-related tasks \cite{b5}. WiFo and WirelessGPT extend the pretraining data to heterogeneous space-time-frequency CSI to support channel prediction under different configurations \cite{b6,b22}. These works show that CSI foundation models have parameter efficiency and transferability in wireless physical-layer tasks.

Despite these advancements, existing channel foundation models still leave room for performance improvement and parameter efficiency. This is because the training processes rarely exploit expert knowledge from wireless channel, such as the sparsity in transformed-domain, and channel multipath parameters, including delays, path powers, and angles. This limits the performance of the CSI foundation model and may increase the training time and model parameters, especially under low-SNR and heterogeneous task settings.
\begin{figure*}[t]
    \centering
    \includegraphics[width=0.95\textwidth]{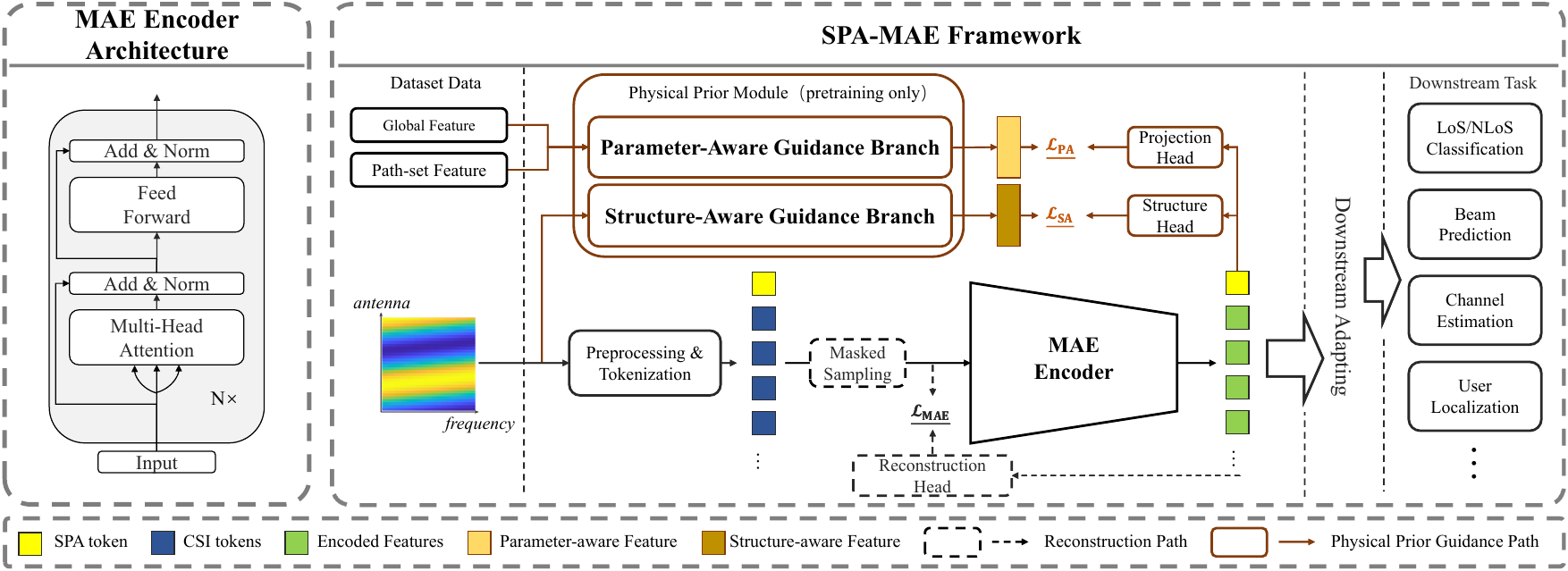}
    \caption{Overview of the proposed SPA-MAE pretraining framework. 
    The left inset illustrates the transformer-based MAE encoder architecture \cite{b3,b9}, and the right panel summarizes the propagation-aware supervision and downstream transfer pipeline.}
    \label{fig:spa_mae_framework}
\end{figure*}

To address these limitations, we propose Structure- and Parameter-Aware Masked Autoencoder (SPA-MAE), a physics-guided CSI foundation model for wireless physical-layer tasks. Specifically, SPA-MAE utilizes a training-only physical prior module to provide two complementary forms of propagation-aware guidance for the MAE encoder. The structure-aware guidance branch guides the model to predict a sparse structure target constructed from the 2D FFT of CSI. The parameter-aware guidance branch uses a pretrained parameter encoder to convert the multipath parameters of each channel sample into an auxiliary target for SPA-MAE pretraining. This training strategy helps SPA-MAE learn CSI features that better reflect the inherent characteristic of wireless channel. After pretraining, the physical prior module is removed, and only the pretrained MAE encoder is retained for downstream adaptation and inference. Experiments on LoS/NLoS classification, beam prediction, channel estimation, and user positioning show that SPA-MAE provides effective transfer performance, outperforms state-of-the-art CSI foundation models with smaller number of model parameters, especially under low-SNR and limited-data conditions.

\section{System Model and Problem Formulation}
\subsection{System Model}

We consider the MISO-OFDM wireless channel and denote the spatial-frequency CSI as
$\mathbf{H}\in\mathbb{C}^{N_a\times N_f}$, where $N_a$ and $N_f$ denote the numbers of antennas and subcarriers, respectively. Under a geometric multipath model, the channel response vector at the $n$-th subcarrier is written as
\begin{equation}
\mathbf{h}[n]=\sum_{p=1}^{P}\alpha_p e^{-j2\pi f_n\tau_p}\mathbf{a}(\theta_p,\phi_p),
\label{eq:geometric_channel}
\end{equation}
where $P$ is the number of propagation paths, $\alpha_p$ and $\tau_p$ denote the complex gain and delay of the $p$-th path, respectively, and $f_n$ is the frequency of the $n$-th subcarrier. The vector $\mathbf{a}(\theta_p,\phi_p)\in\mathbb{C}^{N_a}$ denotes the array response vector corresponding to the elevation angle $\theta_p$ and azimuth angle $\phi_p$ of the $p$-th path. For a general antenna array, it is defined as
\begin{equation}
\begin{aligned}
\mathbf{a}(\theta_p,\phi_p)
&=
\frac{1}{\sqrt{N_a}}
\left[
e^{-j\frac{2\pi}{\lambda}\mathbf{r}_1^{\mathrm T}\mathbf{u}_p},
\ldots,
e^{-j\frac{2\pi}{\lambda}\mathbf{r}_{N_a}^{\mathrm T}\mathbf{u}_p}
\right]^{\mathrm T},\\
\mathbf{u}_p
&=
[
\sin\theta_p\cos\phi_p,\,
\sin\theta_p\sin\phi_p,\,
\cos\theta_p
]^{\mathrm T},
\end{aligned}
\label{eq:array_response}
\end{equation}
where $\lambda$ is the carrier wavelength, $\mathbf{r}_m$ denotes the position vector of the $m$-th antenna element, and $\mathbf{u}_p$ is the unit direction vector of the $p$-th path. By stacking channel response vector at all subcarriers, we obtain the CSI matrix $\mathbf{H}$.
\subsection{Problem Formulation}

Given a CSI sample $\mathbf{H}$, together with training-time guidance from an explicit multipath-parameter set and a transformed-domain structure target, our goal is to learn a lightweight CSI encoder that is transferable across wireless physical-layer tasks. These two guidance signals provide complementary parameter-aware and structure-aware supervision during pretraining. Formally, with trainable parameters $\Theta$, we define the learned encoder as
\begin{equation}
\mathcal{E}_{\Theta}: \mathbf{H} \mapsto \mathbf{F},
\end{equation}
where $\mathbf{F}$ denotes the full encoder output, consisting of the SPA token for global propagation information and encoded CSI token features for fine-grained channel structure. Accordingly, SPA-MAE is pretrained under three complementary constraints, including masked CSI reconstruction loss ($\mathcal{L}_{\mathrm{MAE}}$), the structure-aware prediction loss ($\mathcal{L}_{\mathrm{SA}}$), and the parameter-aware alignment loss ($\mathcal{L}_{\mathrm{PA}}$). Thus, SPA-MAE learns a unified CSI representation that supports diverse downstream tasks beyond masked CSI reconstruction.

\section{Proposed SPA-MAE Framework}
As illustrated in Fig.\,1, this section presents the proposed SPA-MAE framework, including CSI preprocessing and tokenization, the MAE backbone, the physical prior module, and the stage-wise pretraining objectives. The brown components in Fig.\,1 denote the physical prior module and the stage-wise pretraining objectives, which are active only during pretraining and are not used for downstream inference.
\subsection{MAE Backbone with Global SPA Token}

Following the system model in Section II, each CSI sample is denoted by $\mathbf{H}\in\mathbb{C}^{N_a\times N_f}$. To improve training stability, we apply an element-wise compression operator to the CSI magnitude while
preserving the original phase,\footnote{For a nonnegative scalar $a$, the $\mu$-law compression operator is defined as
$\mathcal{C}_{\mu}(a)=\ln(1+\mu a)/\ln(1+\mu)$, where $\mu$ is the
compression coefficient. For a matrix-valued magnitude input,
$\mathcal{C}_{\mu}(\cdot)$ is applied element-wise.} yielding
\begin{equation}
\bar{\mathbf{H}}
=
\mathcal{C}_{\mu}(|\mathbf{H}|)\odot e^{j\angle\mathbf{H}},
\label{eq:mulaw_operator}
\end{equation}
where $\angle\mathbf{H}$ denotes the element-wise phase of $\mathbf{H}$, and $\odot$ denotes element-wise multiplication.

The MAE branch then partitions $\bar{\mathbf{H}}$ into $K=N_aN_f/L$ non-overlapping CSI patches, each containing $L$ complex entries. Let $\bar{\mathbf{H}}_k\in\mathbb{C}^{L}$ denote the $k$-th vectorized patch. The real-valued token and the resulting token sequence are defined as
\begin{equation}
\begin{aligned}
\mathbf{x}_k
&=[\Re(\bar{\mathbf{H}}_k),\Im(\bar{\mathbf{H}}_k)]^\mathrm{T}
\in\mathbb{R}^{2L},\quad k=1,\ldots,K,\\
\mathbf{X}
&=[\mathbf{x}_1,\ldots,\mathbf{x}_K]\in\mathbb{R}^{K\times2L}.
\end{aligned}
\label{eq:csi_tokenization}
\end{equation}

Given the CSI token sequence $\mathbf{X}$, the MAE backbone first embeds each token while preserving its position implied by the row-wise token order. Specifically, under this order, the $k$-th token is assigned a two-dimensional positional label $(r_k,s_k)$, where $r_k$ denotes the antenna-row position and $s_k$ denotes the frequency-segment position within that row. Each token is then mapped to the encoder dimension by a multi-layer perceptron (MLP) and added with a CSI-aware positional embedding as
\begin{equation}
\tilde{\mathbf{z}}_k
=
f_{\mathrm{in}}(\mathbf{x}_k)
+
\mathbf{e}_{\mathrm{row}}(r_k)
+
\mathbf{e}_{\mathrm{seg}}(s_k),
\quad k=1,\ldots,K,
\end{equation}
where $f_{\mathrm{in}}(\cdot)$ is a learnable MLP, $\mathbf{e}_{\mathrm{row}}(\cdot)$ is a learnable row embedding initialized by sinusoidal encoding, and $\mathbf{e}_{\mathrm{seg}}(\cdot)$ is a learnable segment embedding shared across patches with the same intra-row segment index.

A learnable SPA token $\tilde{\mathbf{z}}_0\in\mathbb{R}^{d}$ is prepended to the CSI token sequence to aggregate global information, where $d$ denotes the encoder embedding dimension. The resulting token sequence is
\begin{equation}
\tilde{\mathbf{Z}}
=
[\tilde{\mathbf{z}}_0;\tilde{\mathbf{z}}_1;\ldots;\tilde{\mathbf{z}}_K]
\in\mathbb{R}^{(K+1)\times d}.
\end{equation}
During pretraining, random masking samples indices only from the CSI-token set $\{1,\ldots,K\}$. The SPA token is always kept. The encoder takes the SPA token and visible CSI tokens as input, and the decoder reconstructs the masked CSI tokens using mask tokens and restoration indices.

For downstream transfer, masking is disabled, and all CSI tokens are fed into the encoder together with the SPA token. The retained encoder outputs the encoded token sequence
\begin{equation}
\mathbf{F}
=
[\mathbf{z}_0;\mathbf{z}_1;\ldots;\mathbf{z}_K]
\in
\mathbb{R}^{(K+1)\times d}.
\end{equation}

\subsection{Physical Prior Module}
\label{subsec:physical_prior}

During pretraining, the physical prior module exploits the geometric multipath nature of wireless channels from two complementary views. In the spatial-frequency CSI, multiple paths are superposed and their propagation structure is not explicitly separated. To expose such structure, SPA-MAE uses two sources of propagation-aware supervision, namely structure-aware guidance from a transformed-domain CSI target and parameter-aware guidance from explicit multipath priors. This subsection defines the two guidance signals, and the following subsection introduces their stage-wise optimization objectives.

\subsubsection{Structure-Aware Guidance Branch}

Complementary to the parameter-aware branch, the structure-aware branch derives structural supervision directly from CSI. Although multipath components are superposed in the spatial-frequency domain, a 2D FFT maps the CSI into a transformed domain where dominant components become more separable and sparse. This motivates the construction of a transformed-domain structure target to regularize the MAE encoder during pretraining.

Specifically, we apply a 2D FFT to the input CSI, take the magnitude, and compress it with the same $\mu$-law operator used in Sec.~III-A. The structure target is then obtained by vectorization as
\begin{equation}
\mathbf{s}
=
\mathrm{vec}\!\left(
\mathcal{C}_{\mu}\left(|\mathrm{FFT2}(\mathbf{H})|\right)
\right)
\in\mathbb{R}^{N_aN_f},
\end{equation}
where $\mathrm{FFT2}(\cdot)$ denotes the 2D FFT and $\mathrm{vec}(\cdot)$ vectorizes the transformed-domain magnitude map. The magnitude is used because the transformed-domain structure is mainly reflected by the energy distribution of dominant components, while the $\mu$-law compression stabilizes the dynamic range of the target.

During pretraining, a compressed-sensing-inspired structure head attached to the  SPA token output predicts $\hat{\mathbf{s}}$. This head first projects the  SPA token representation to a hidden space, reconstructs the high-dimensional structure target through a linear reconstruction layer, and then applies learnable soft-thresholding followed by a non-negative output constraint to encourage sparse dominant components. The prediction $\hat{\mathbf{s}}$ is later aligned with $\mathbf{s}$ through the structure-aware alignment objective. In this way, the MAE encoder is encouraged to preserve not only reconstructable local CSI content but also global transformed-domain structure information.

\subsubsection{Parameter-Aware Guidance Branch}

\begin{figure}[!t]
    \centering
    \includegraphics[width=0.90\columnwidth]{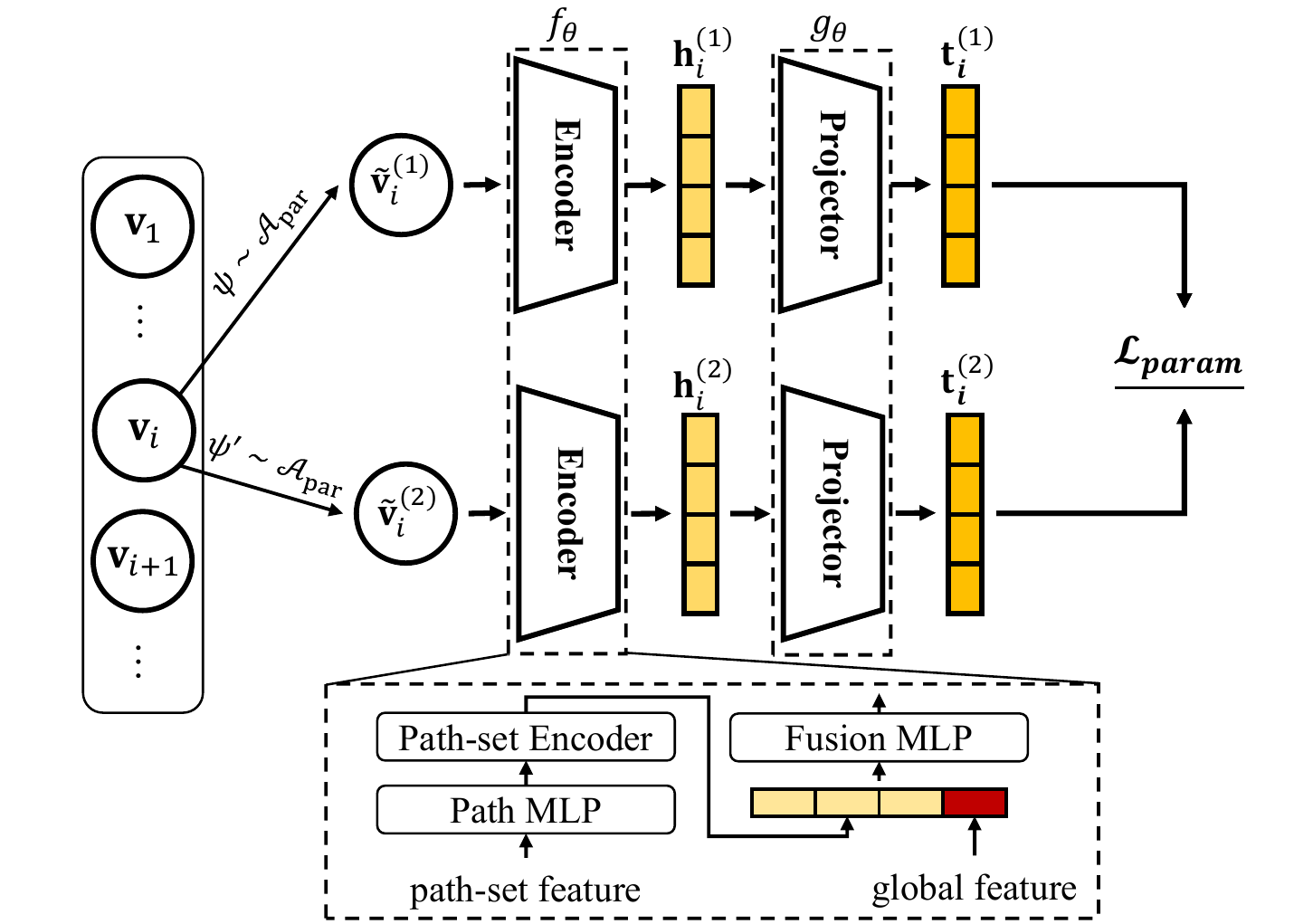}
    \caption{Parameter-Aware Guidance Branch. Multipath priors
    are encoded into parameter-aware targets for SPA token alignment
    after dual-view contrastive pretraining.}
    \label{fig:teacher_model}
\end{figure}

As shown in Fig.\,2, this branch encodes explicit multipath priors into parameter-aware targets for SPA-MAE pretraining. For sample $i$, each valid path provides delay, path power, elevation, and azimuth. Following the direction-vector definition in \eqref{eq:array_response}, these parameters are converted into a path-wise physical descriptor
\begin{equation}
\mathbf{q}_{i,p}
=
\big[\Delta\tau_{i,p},\ \rho_{i,p},\ \mathbf{u}_{i,p}^{\top}\big]^{\top},
\label{eq:path_descriptor}
\end{equation}
where $p$ denotes the path index, $\Delta\tau_{i,p}$ is the delay relative to the earliest path of sample $i$, $\rho_{i,p}$ is the path power, and $\mathbf{u}_{i,p}$ is obtained from (2) using the
elevation-azimuth pair $(\theta_{i,p},\phi_{i,p})$. In addition to these path-wise descriptors, the path loss $\eta_i$ provided by DeepMIMO is used as a sample-level global feature $\mathbf{g}_i=[\eta_i]^{\top}$ to characterize the overall link attenuation.

Since different CSI samples may contain different numbers of valid paths, each set $\mathcal{Q}_i$ is zero-padded to $N_p$ path slots to form a fixed-length path-parameter input. The parameter-aware encoder takes $\mathbf{v}_i = (\mathcal{Q}_i, \mathbf{g}_i)$ as input. The resulting path embeddings are then fed into a path-set encoder, which performs set encoding and attention-based aggregation to produce a path-set feature. This path-set feature is fused with the path-loss global feature $\mathbf{g}_i$ by a fusion MLP. The encoder output is denoted as $\mathbf{h}_i=f_\theta(\mathbf{v}_i)$, and a projector $g_\theta$ further maps it to a normalized parameter-aware target $\mathbf{t}_i=g_\theta(\mathbf{h}_i)$.

Before being used for parameter-aware alignment, the parameter-aware encoder is pretrained by dual-view contrastive learning on physical-parameter inputs. Two augmentation operators $\psi,\psi'\sim\mathcal{A}_{\mathrm{par}}$ are independently sampled to generate two correlated views of the same input, i.e., $\tilde{\mathbf{v}}_i^{(1)}=\psi(\mathbf{v}_i)$ and $\tilde{\mathbf{v}}_i^{(2)}=\psi'(\mathbf{v}_i)$. Here, $\mathcal{A}_{\mathrm{par}}$ includes path permutation, path dropping or mask perturbation, parameter perturbation, and global-feature dropout. The two views are passed through the shared encoder $f_\theta$ and projector $g_\theta$ to produce $\mathbf{t}_i^{(1)}$ and $\mathbf{t}_i^{(2)}$. As illustrated in Fig.\,2, the two view embeddings of the same sample, $\mathbf{t}_i^{(1)}$ and $\mathbf{t}_i^{(2)}$, are treated as a positive pair. We re-index the two-view embeddings $\{\mathbf{t}_i^{(1)},\mathbf{t}_i^{(2)}\}_{i=1}^{B}$ as $\{\mathbf{t}_r\}_{r=1}^{2B}$, and let $\mathbf{t}_r^{+}$ denote the paired embedding from the other view of the same sample. The parameter-aware contrastive loss is defined as
\begin{equation}
\mathcal{L}_{\mathrm{param}} = -
\frac{1}{2B}
\sum_{r=1}^{2B}
\log
\frac{\exp(\mathrm{sim}(\mathbf{t}_r,\mathbf{t}_r^+)/\kappa_{\mathrm{par}})}
{\sum_{u\ne r}\exp(\mathrm{sim}(\mathbf{t}_r,\mathbf{t}_u)/\kappa_{\mathrm{par}})},
\end{equation}
where $B$ is the batch size, $\kappa_{\mathrm{par}}$ is the contrastive temperature, and $\mathrm{sim}(\cdot,\cdot)$ denotes cosine similarity. After this contrastive pretraining, the parameter-aware encoder is fixed and used to provide parameter-aware targets for subsequent SPA token alignment.

\subsection{Stage-Wise Training Objectives}
\label{subsec:optimization}

After the parameter-aware encoder is pretrained and fixed, the MAE encoder is trained in two stages, with masked CSI reconstruction kept as the primary objective throughout training. Stage I introduces structure-guided pretraining on top of masked reconstruction, while Stage II further adds parameter-aware alignment with the fixed parameter-aware encoder. After pretraining, the physical prior module and auxiliary heads are discarded, and only the MAE encoder is retained for downstream adaptation and inference.

\subsubsection{Stage I: Reconstruction and Structure-Guided Pretraining}

Using the  SPA token output, the structure head predicts
$\hat{\mathbf{s}}\in\mathbb{R}^{N_aN_f}$. The masked reconstruction loss and
the structure-aware prediction loss are defined as
\begin{equation}
\mathcal{L}_{\mathrm{MAE}}=
\frac{1}{|\mathcal{M}|}
\sum_{k\in\mathcal{M}}
\left\|
\frac{\mathbf{x}_k-\hat{\mathbf{x}}_k}{\sigma_{\mathbf{X}}}
\right\|^2,
\end{equation}
\begin{equation}
\mathcal{L}_{\mathrm{SA}}=
\frac{1}{N_aN_f}
\left\|
\frac{\hat{\mathbf{s}}-\mathbf{s}}{\sigma_{\mathbf{s}}}
\right\|^2,
\end{equation}
where $\mathcal{M}$ denotes the masked-token index set, $\hat{\mathbf{x}}_k$
is the reconstructed CSI token, and $\sigma_{\mathbf{X}}$ and
$\sigma_{\mathbf{s}}$ are the standard deviations of the CSI tokens and
structure target, respectively.

\subsubsection{Stage II: Parameter-Aware Alignment and Final Objective}
For sample $i$, the fixed parameter-aware encoder provides the target $\mathbf{t}_i$, while a projection head maps the SPA token output $\mathbf{z}_{0,i}$ to a normalized representation $\mathbf{r}_i$. Let $\mathbf{T}=[\mathbf{t}_1;\ldots;\mathbf{t}_B]$ and $\mathbf{R}=[\mathbf{r}_1;\ldots;\mathbf{r}_B]$ denote the batch-level parameter-aware targets and projected SPA token representations, respectively, where $B$ is the batch size. To incorporate propagation information from multipath parameters, we align $\mathbf{R}$ with $\mathbf{T}$ through the following relation alignment and contrastive matching losses
\begin{equation}
\mathcal{L}_{\mathrm{rel}} =
D_{\mathrm{KL}}\left(
\mathrm{softmax}(\mathbf{T}\mathbf{T}^{\top}/\kappa_{\mathrm{pa}})
\middle\|
\mathrm{softmax}(\mathbf{R}\mathbf{R}^{\top}/\kappa_{\mathrm{pa}})
\right),
\end{equation}
\begin{equation}
\mathcal{L}_{\mathrm{con}} =
-\frac{1}{B}\sum_{b=1}^{B}
\log
\frac{\exp(\mathbf{r}_b^{\top}\mathbf{t}_b/\kappa_{\mathrm{pa}})}
{\sum_{j=1}^{B}\exp(\mathbf{r}_b^{\top}\mathbf{t}_j/\kappa_{\mathrm{pa}})},
\end{equation}
\begin{equation}
\mathcal{L}_{\mathrm{PA}}=
\alpha \mathcal{L}_{\mathrm{rel}}+\beta \mathcal{L}_{\mathrm{con}}.
\end{equation}
Here, $D_{\mathrm{KL}}(\cdot\|\cdot)$ denotes the KL divergence, $\kappa_{\mathrm{pa}}$ is the distillation temperature, and $\alpha$ and $\beta$ are the weights of relation alignment and contrastive matching, respectively. The final Stage~II objective is
\begin{equation}
\mathcal{L}_{\mathrm{train}}=
\lambda_{\mathrm{mae}}\mathcal{L}_{\mathrm{MAE}}
+\lambda_{\mathrm{sa}}\mathcal{L}_{\mathrm{SA}}
+\lambda_{\mathrm{pa}} \mathcal{L}_{\mathrm{PA}},
\end{equation}
where $\lambda_{\mathrm{mae}}$, $\lambda_{\mathrm{sa}}$, and $\lambda_{\mathrm{pa}}$ are the weights of the three training objectives.

\begin{figure}[!t]
    \centering

    \begin{subfigure}{\columnwidth}
        \centering
        \includegraphics[width=0.85\columnwidth,trim=10mm 0mm 0mm 0mm,clip]{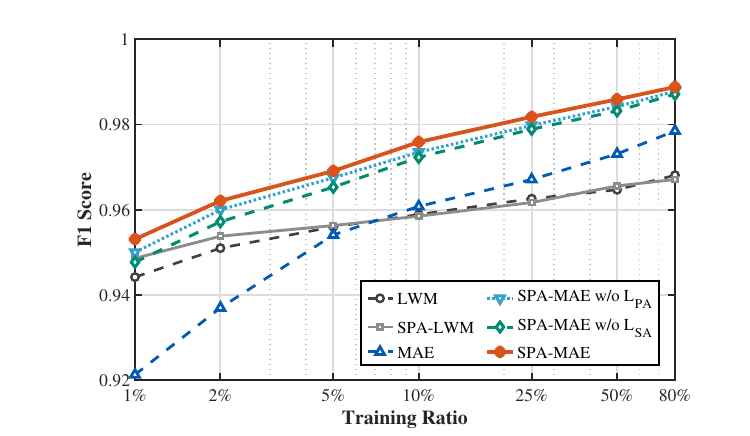}
        \caption{SNR=$20$ dB.}
        \label{fig:overall_results_snr20}
    \end{subfigure}

    \vspace{0.5mm}

    \begin{subfigure}{\columnwidth}
        \centering
        \includegraphics[width=0.85\columnwidth,trim=10mm 0mm 0mm 0mm,clip]{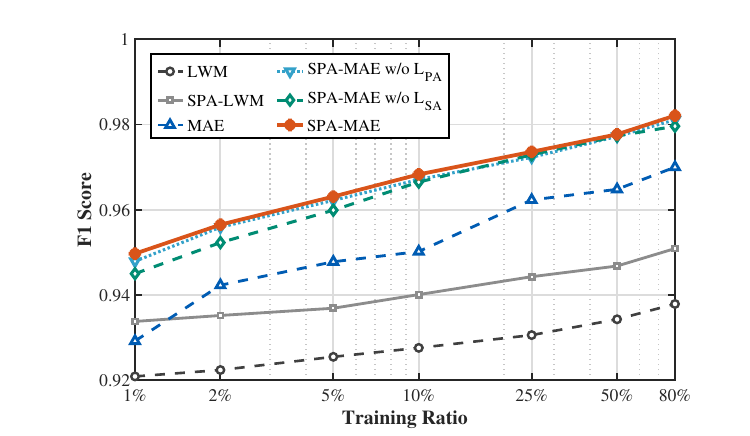}
        \caption{SNR=$0$ dB.}
        \label{fig:overall_results_snr0}
    \end{subfigure}

    \caption{Overall comparison and ablation study on LoS/NLoS classification. 
    (a) F1 score comparison under varying training ratios at SNR=$20$ dB. 
    (b) F1 score comparison under varying training ratios at SNR=$0$ dB. 
    In (a) and (b), ``SPA-MAE w/o $\mathcal{L}_{\mathrm{SA}}$'' and 
    ``SPA-MAE w/o $\mathcal{L}_{\mathrm{PA}}$'' correspond to removing 
    $\mathcal{L}_{\mathrm{SA}}$ and $\mathcal{L}_{\mathrm{PA}}$, respectively.}
    \label{fig:overall_results}
\end{figure}
\section{Performance Evaluation}

\subsection{Experimental Setup}
Unless otherwise specified, all optimization uses AdamW
with batch size 1024 on seen DeepMIMO~\cite{b2} scenarios,
covering \textit{O1}, \textit{Boston5G}, \textit{ASU campus1}, and multiple city
scenarios. The MAE encoder operates on CSI inputs of size
$N_a \times N_f = 32 \times 32$ with patch length $L=16$,
masking ratio $r=0.5$, and embedding dimension $d=64$.
During pretraining, 60\% of CSI samples are corrupted by
complex Gaussian noise with SNR uniformly sampled from
$[-20,20]$ dB. The parameter-aware encoder is pretrained for
60 epochs with $\mathcal{L}_{\mathrm{param}}$ temperature
$\kappa_{\mathrm{par}}=0.07$. The MAE encoder is then
trained for 120 epochs with reconstruction and structure
guidance using $\lambda_{\mathrm{mae}}=1.0$ and
$\lambda_{\mathrm{sa}}=0.2$, followed by 100 epochs of
parameter-aware alignment with $\lambda_{\mathrm{mae}}=1.0$,
$\lambda_{\mathrm{sa}}=0.05$, $\lambda_{\mathrm{pa}}=0.1$,
$\alpha=0.7$, $\beta=0.3$, and $\kappa_{\mathrm{pa}}=0.5$.
After pretraining, the physical prior module is removed, and
only the MAE encoder is retained for downstream tasks.

We evaluate SPA-MAE on LoS/NLoS classification and further assess its transferability on beam prediction, channel estimation, and user positioning. All methods are tested on the same dataset using six unseen DeepMIMO~\cite{b2} scenarios, including \textit{Denver}, \textit{Fort Worth}, \textit{Oklahoma}, \textit{Indianapolis}, \textit{Santa Clara}, and \textit{San Diego}. Unless otherwise specified, all compared pretrained representations use identical downstream heads within each task, and all backbone encoders remain frozen during downstream adaptation.

\subsection{Overall Comparison and Ablation Study}

We compare the proposed SPA-MAE with the following baselines:
\begin{itemize}
    \item \textbf{MAE}: The same backbone as SPA-MAE,
    pretrained only by masked CSI reconstruction without the
    physical prior module.
    \item \textbf{LWM}~\cite{b1}: A transformer-based wireless
    channel foundation model that is self-supervised with masked
    channel modeling to generate task-agnostic CLS and channel
    embeddings for downstream wireless tasks.
    \item \textbf{SPA-LWM}: An enhanced LWM variant equipped
    with the proposed propagation-aware supervision, used to
    verify the portability of the physical-prior guidance beyond
    the MAE backbone.
\end{itemize}

For downstream adaptation, the encoded SPA token $z_0$ is used for tasks that rely on global channel information, including LoS/NLoS classification and user positioning, while the outputted feature $\mathbf{F}$ is used for beam prediction and channel estimation. Similarly, LWM uses its CLS embedding and channel embedding for the same two roles. The retained SPA-MAE encoder contains about 0.33M parameters, whereas the retained LWM encoder contains about 0.60M parameters.

As shown in Fig.\,3, SPA-MAE achieves
the best overall F1 performance under both 20 dB and 0 dB
across all training ratios, confirming the effectiveness of the
proposed pretraining architecture on the MAE backbone.
SPA-LWM also improves over LWM, especially under low SNR
and limited data.

Fig.\,3 further shows the effect of the two
propagation-aware losses. Compared with MAE, the variants with either
$\mathcal{L}_{\mathrm{SA}}$ or $\mathcal{L}_{\mathrm{PA}}$ achieve
higher F1 scores, and the full SPA-MAE obtains the best performance.
This indicates that the structure-aware and parameter-aware supervision
bring complementary improvements. Since the physical prior module is
removed after pretraining, these gains are achieved using only the
retained MAE encoder during inference.
\subsection{Downstream Task Evaluation}
\label{subsec:downstream}
We next evaluate transferability on channel estimation, beam prediction, and user positioning. LWM remains the main wireless-native baseline for full-representation transfer tasks.

\begin{figure}[!t]
    \centering
    \includegraphics[width=0.85\columnwidth,trim=10mm 0mm 0mm 0mm,clip]{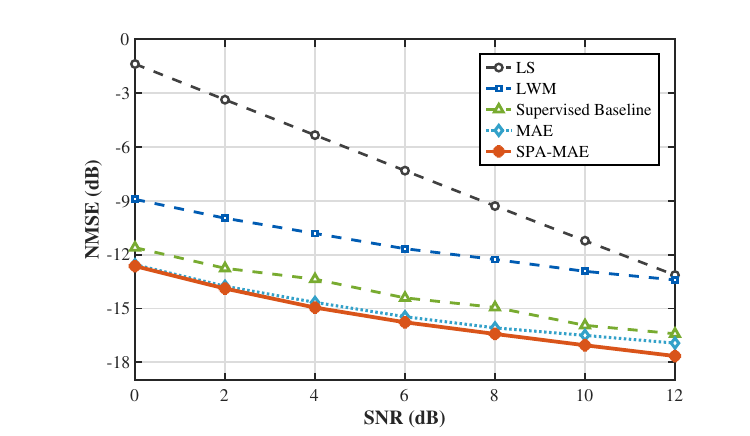}
    \caption{NMSE comparisons on channel estimation task.}
    \label{fig:channel_estimation}
\end{figure}

\textbf{Channel Estimation:} For channel estimation, receiver-side observations constructed from the received signals and known pilots are tokenized and fed into the pretrained backbone. The resulting outputted features are sent to a lightweight transformer-based predictor to reconstruct the complete CSI matrix. The proposed SPA-MAE is compared with the conventional LS estimator, LWM, MAE, and a supervised baseline. The supervised baseline uses the same backbone architecture and parameter budget as SPA-MAE's MAE backbone but is trained from scratch without pretraining. In contrast, MAE uses the same backbone but is pretrained only with masked CSI reconstruction, serving as a backbone-matched baseline to isolate the effect of the proposed propagation-aware pretraining. As shown in Fig.\,4, SPA-MAE consistently achieves the lowest NMSE across the evaluated SNRs. The gap between SPA-MAE and MAE highlights the benefit of the proposed propagation-aware pretraining, while the overall results remain competitive with the supervised baseline. A possible reason for the weaker performance of LWM is that its BERT-style architecture is more oriented toward global contextual modeling and may be less suitable for reconstruction-heavy channel estimation.

\begin{figure}[!t]
    \centering
    \includegraphics[width=0.92\columnwidth]{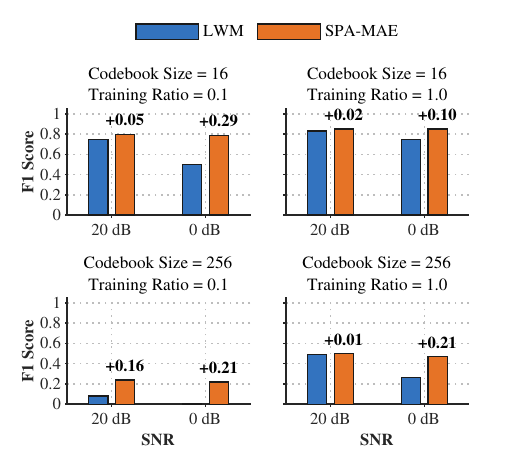}
    \caption{Beam prediction performance under different SNR levels, codebook sizes, and training ratios.}
    \label{fig:beam_prediction}
\end{figure}

\textbf{Beam Prediction:} For beam prediction, we perform cross-frequency prediction by using the representations extracted from the $3.5$ GHz CSI to predict the best beam index at $28$ GHz under different codebook sizes, training ratios, and SNR conditions. The outputted features of SPA-MAE and the channel embedding of LWM are fed into the same residual 1D-CNN predictor~\cite{b21}, and the performance is measured by the F1 score. As shown in Fig.\,5, where each subplot fixes the codebook size and training ratio and annotates the F1-score gain over LWM, SPA-MAE outperforms LWM in all evaluated settings. Consistent with the low-SNR and limited-data trends in Fig.\,5, the gain is more evident at low SNR and with limited training data, and further grows with larger codebooks. This indicates that the proposed pretraining architecture improves representation transfer for challenging cross-frequency beam selection settings.

\textbf{User Positioning:} For user positioning, the SPA token of SPA-MAE and the CLS token of LWM are fed to the same simple MLP regressor to predict the 2D relative user location, and performance is evaluated by the mean distance error (MDE). As shown in Table\,I, SPA-MAE consistently achieves the lowest positioning error across all evaluated scenarios. This suggests that the encoded SPA token provides effective global spatial cues for positioning.

\section{Conclusion}
This paper proposed SPA-MAE, a physics-guided CSI foundation model for wireless physical layer tasks. By introducing parameter-aware guidance and structure-aware guidance during pretraining, SPA-MAE enhances the MAE encoder without adding inference-complexity. After training, the entire physical prior module is removed, and only the retained MAE encoder is used for downstream tasks. Experimental results on LoS/NLoS classification, beam prediction, channel estimation, and user positioning show that this retained encoder provides competitive transfer performance and favorable parameter efficiency relative to representative baselines.
\begin{table}[!t]
\centering
\caption{Performance comparison on the user positioning task in terms of mean distance error (m).}
\label{tab:user_localization}
\setlength{\tabcolsep}{4pt}
\renewcommand{\arraystretch}{1.15}
\resizebox{\columnwidth}{!}{%
\begin{tabular}{lcccccc}
\toprule
\textbf{Method} & \textbf{Denver} & \textbf{Indianapolis} & \textbf{Oklahoma} & \textbf{Fort Worth} & \textbf{Santa Clara} & \textbf{San Diego} \\
\midrule
\textbf{LWM}     & 28.05 & 15.06 & 16.67 & 21.89 & 15.63 & 17.35 \\
\textbf{SPA-MAE} & 12.37 & 5.89  & 6.62  & 6.11  & 6.64  & 6.90  \\
\bottomrule
\end{tabular}%
}
\end{table}


\begin{thebibliography}{00}
\bibitem{b11} K. B. Letaief, W. Chen, Y. Shi, J. Zhang, and Y.-J. A. Zhang, ``The roadmap to 6G: AI empowered wireless networks,'' \emph{IEEE Commun. Mag.}, vol. 57, no. 8, pp. 84--90, Aug. 2019.

\bibitem{b12} W. Saad, M. Bennis, and M. Chen, ``A vision of 6G wireless systems: Applications, trends, technologies, and open research problems,'' \emph{IEEE Netw.}, vol. 34, no. 3, pp. 134--142, May/Jun. 2020.

\bibitem{b15} W. Jin, J. Zhang, C.-K. Wen, S. Jin, X. Li, and S. Han, ``Low-complexity joint beamforming for RIS-assisted MU-MISO systems based on model-driven deep learning,'' \emph{IEEE Trans. Wireless Commun.}, vol. 23, no. 7, pp. 6968--6982, Jul. 2024.

\bibitem{b16} M. Alrabeiah and A. Alkhateeb, ``Deep learning for mmWave beam and blockage prediction using sub-6 GHz channels,'' \emph{IEEE Trans. Commun.}, vol. 68, no. 9, pp. 5504--5518, Sep. 2020.

\bibitem{b20} H. He, S. Jin, C.-K. Wen, F. Gao, G. Y. Li, and Z. Xu, ``Model-driven deep learning for physical layer communication'', \emph{IEEE Wireless Communications}, vol. 26, no. 5, pp. 77-83, Oct. 2019. 

\bibitem{b17} J. Devlin, M.-W. Chang, K. Lee, and K. Toutanova, ``BERT: Pre-training of deep bidirectional transformers for language understanding,'' in \emph{Proc. North Amer. Chapter Assoc. Comput. Linguistics: Human Language Technologies (NAACL-HLT)}, Minneapolis, MN, USA, Jun. 2019, pp. 4171--4186.

\bibitem{b18} H. Bao, L. Dong, S. Piao, and F. Wei, ``BEiT: BERT pre-training of image transformers,'' in \emph{Proc. Int. Conf. Learn. Represent. (ICLR)}, Virtual, Apr. 2022.

\bibitem{b19} A. Kirillov, E. Mintun, N. Ravi, H. Mao, C. Rolland, L. Gustafson, T. Xiao, S. Whitehead, A. C. Berg, W.-Y. Lo, P. Doll\'{a}r, and R. Girshick, ``Segment anything,'' in \emph{Proc. IEEE/CVF Int. Conf. Comput. Vis. (ICCV)}, Paris, France, Oct. 2023, pp. 4015--4026.

\bibitem{b4} J. Guo, P. Jiang, C.-K. Wen, S. Jin, and J. Zhang, ``LVM4CSI: Enabling direct application of pre-trained large vision models for wireless channel tasks,'' arXiv preprint arXiv:2507.05121, 2025.

\bibitem{b10} X. Liu, S. Gao, B. Liu, X. Cheng, and L. Yang, ``LLM4WM: Adapting LLM for wireless multi-tasking,'' \emph{IEEE Trans. Mach. Learn. Commun. Netw.}, vol. 3, pp. 835--847, Jul. 2025.

\bibitem{b1} S. Alikhani, G. Charan, and A. Alkhateeb, ``Large wireless model (LWM): A foundation model for wireless channels,'' arXiv preprint arXiv:2411.08872, 2024.

\bibitem{b5} J. Jiang, X. Ruan, and S. Xu, ``CSI-MAE: A masked autoencoder-based channel foundation model,'' arXiv preprint arXiv:2601.03789, 2026.

\bibitem{b6} B. Liu, S. Gao, X. Liu, X. Cheng, and L. Yang, ``WiFo: Wireless foundation model for channel prediction,'' \emph{Sci. China Inf. Sci.}, vol. 68, no. 6, Art. no. 162302, Jun. 2025.

\bibitem{b22} T. Yang, P. Zhang, M. Zheng, Y. Shi, L. Jing, J. Huang, and N. Li, ``WirelessGPT: A generative foundation model for multi-task integrated sensing and communication,'' \emph{IEEE J. Sel. Areas Commun.}, vol. 44, pp. 2259--2273, 2026.

\bibitem{b9} A. Vaswani, N. Shazeer, N. Parmar, J. Uszkoreit, L. Jones, A. N. Gomez, {\L}. Kaiser, and I. Polosukhin, ``Attention is all you need,'' in \emph{Proc. Adv. Neural Inf. Process. Syst. (NeurIPS)}, 2017, pp. 5998--6008.

\bibitem{b3} K. He, X. Chen, S. Xie, Y. Li, P. Doll\'{a}r, and R. Girshick, ``Masked autoencoders are scalable vision learners,'' in \emph{Proc. IEEE/CVF Conf. Comput. Vis. Pattern Recognit.}, 2022, pp. 16000--16009.

\bibitem{b2} A. Alkhateeb, ``DeepMIMO: A generic deep learning dataset for millimeter wave and massive MIMO applications,'' in \emph{Proc. Inf. Theory Appl. Workshop (ITA)}, San Diego, CA, USA, Feb. 2019, pp. 1--8.

\bibitem{b21} K. He, X. Zhang, S. Ren, and J. Sun, ``Deep residual learning for image recognition,'' in \emph{Proc. IEEE Conf. Comput. Vis. Pattern Recognit. (CVPR)}, Las Vegas, NV, USA, Jun. 2016, pp. 770--778.
\end{thebibliography}
\end{document}